\documentclass[aps,a4paper,twocolumn,showpacs]{revtex4}

\usepackage{graphicx}
\usepackage{dcolumn}
\usepackage{amsmath}
\usepackage{epsfig}
\usepackage{pifont}
\usepackage{amssymb}

\begin{document}

\title{Reversing a granular flow on a vibratory conveyor}

\author{R.\ Grochowski}
\author{P.\ Walzel}

\affiliation{\it Mechanische Verfahrenstechnik, Universit\"at Dortmund,
D--44227 Dortmund, Germany}

\author{M. Rouijaa}
\author{C.A.\ Kruelle\footnote{Electronic mail:
christof.kruelle@uni-bayreuth.de}}
\author{I.\ Rehberg}

\affiliation{\it Experimentalphysik V, Universit\"at Bayreuth, D--95440 Bayreuth,
Germany}

\date{November 25, 2003}

\begin{abstract}
Experimental results are presented on the transport properties of granular
materials on a vibratory conveyor. For circular oscillations of the shaking
trough a non-monotonous dependence of the transport velocity on the
normalized acceleration $\Gamma$ is observed. Two maxima are separated
by a regime, where the granular flow is much slower and, in a certain driving
range, even {\em reverses} its direction. A similar behavior is found for a
single solid body with a low coefficient of restitution, whereas an individual
glass bead of 1 mm diameter is propagated in the same direction for all
accelerations.
\end{abstract}


\maketitle

The controlled transport of bulk cargoes by means of vibratory conveyors is
of major importance for a whole variety of industrial processes
\cite{pajer88,rademacher94,sloot96}. The granular material is usually (i)
agitated by a stick-slip drag on a horizontally vibrated deck with asymmetric
forward and backward motions, (ii) forced to perform ballistic flights if the
vertical component of the acceleration exceeds gravity, or (iii) can be
transported horizontally by a vertically oscillating asymmetric
sawtooth-shaped profile of the base \cite{derenyi98,farkas99}.

Since these transport phenomena involve the nonlinear interaction of
many-particle systems with complex behavior leading to self-organized
spatiotemporal patterns, the investigation of vibrating granular materials has
become a challenging subject to physicists, too \cite{jaeger96,kadanoff99}.
Such intriguing phenomena as surface waves
\cite{douady89,pak93,melo94,metcalf97}, organized clusters
\cite{strassburger00} or segregation effects
\cite{aumaitre01,aumaitre03,moon03}
have attracted a lot of attention.

In this letter, we introduce a conveyor system with convertible modes of
oscillation based on the combined forces of four rotating unbalanced masses.
First results are reported on the surprising transport properties caused by
{\em circular} vibrations of the trough.

\begin{figure}[tb]
\noindent
\begin{minipage}{74mm}
\epsfxsize=74mm
\epsfbox{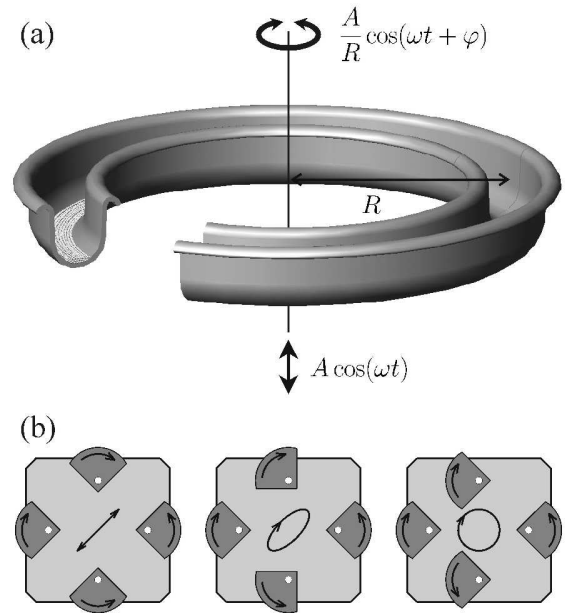}
\caption{(a) Sketch of the circular vibratory conveyor. By adjusting the phase
shift $\varphi$ between the vertical and the torsional vibration each part of the
annular trough can be forced on a specified trajectory. (b) Schematic side
views of a driving unit with four rotating unbalanced masses, for three
principal modes of oscillation: linear ($\varphi = 0$), elliptical
($\varphi = \frac{\pi}{4}$), and circular ($\varphi = \frac{\pi}{2}$).}
\label{setup}
\end{minipage}
\end{figure}

The circular vibratory conveyor, shown schematically in Fig.\ 1(a), is a
prototype apparatus specially developed to investigate the transport under
principal oscillation modes, i.e.\ linear, elliptical, and circular for a long
running time, without disturbing boundary conditions.  In our experiments,
the transporting trough has the form of a horizontally oriented ring with
radius $R$ = 22.5 cm and width $w$ = 5 cm and is suspended on adjustable
columns via elastic bands.  By means of four driving units, symmetrically
positioned below the trough, the conveyor can vibrate with defined amplitude
and oscillation pattern along its entire circumference.  This motion can be
described by a trajectory performed on a cylindrical surface, consisting of a
vertical oscillation $z(t) = A\cos(\omega t)$ superposed with a torsional
vibration $\phi(t) = A/R \cos(\omega t+\varphi)$ around the symmetry axis of
the apparatus, where $\varphi$ is the fixed phase shift between the two
oscillations. Note that, since the ratio $A/R$ is only about 1 percent, the path
of each segment of the tray can be considered to lie almost in a vertical plane.
If, for example, the phase shift $\varphi$ is chosen to be $\pi/2$, then each
point on the trough traces a circular path in a vertical plane tangent to the
trough at that point.

For achieving this kind of motion with the inherent possibility to generate
different modes of oscillation a special adjustable drive is acquired.
Unbalanced-mass vibrators are well established in industrial applications
since long time. Their working principle is based on a centrifugal force
$F = m_{\rm u} r_{\rm u}(2\pi f)^{2}$ produced by an unbalanced mass
$m_{\rm u}$ rotating with frequency $f$, with $r_{\rm u}$ being the distance
between the center of gravity of the unbalanced (eccentric) mass and its
rotation axis. A motor driving a load $M$ with one single unbalanced mass
will create, for frequencies well above resonance, a circular vibration with an
amplitude $A_{\infty} = r_{\rm u}m_{\rm u}/M$. A linear motion can be excited
by the joint action of two equal vibrators rotating in opposite directions. In
consequence, by combining two such linear vibrators oriented
perpendicularly to each other, it is possible to generate any desired Lissajous
figure by adjusting the phase shift $\varphi$ between the two oscillations.
Three examples of possible modes are shown in Fig.\ 1(b).

\begin{figure}[tb]
\noindent
\begin{minipage}{74mm}
\epsfxsize=74mm
\epsfbox{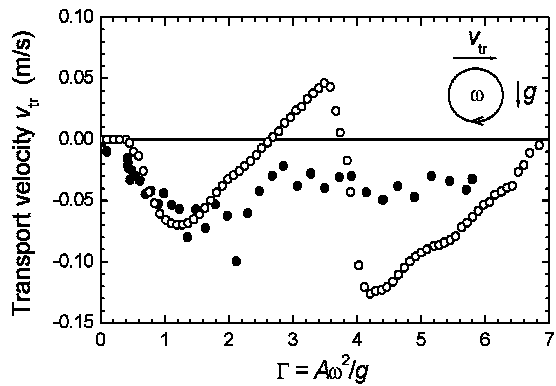}
\caption{Transport velocity $v_{\rm tr}(\Gamma)$ of a granular flow
($\approx$ 300 000 glass beads with 1 mm diameter, $\circ$) on the vibratory
conveyor, compared with the mean velocity of one single glass bead
($\bullet$). Conveyor amplitude $A_{\infty} = 1.7$ mm.}
\label{granulate}
\end{minipage}
\end{figure}

Each driving unit is built as the above described system of four unbalanced
masses, which are placed vertically on both sides of the unit. The vibrators
are fixed to the carrying plates by bolts located in circular grooves, which
enable the adjustment of the vibration angle $\alpha$ between $8^{\circ}$ and
$82^{\circ}$. The oscillation amplitude can be adjusted in steps by changing
the number of impaled unbalanced masses.

The four driving units are coupled to the motor via a common central gear
box, which keeps all drives in the same phase. The driving torque is
transmitted from the gear box to the vibrators by use of rotating rods
connected with compensation clutches. The conveyor is driven by an electric
motor (Siemens Combimaster 1UA7, with integrated frequency inverter). By
changing its rotation frequency $f$, the number of unbalanced masses
$m_{\rm u}$, and their angular alignment,  the adjustment of all required
oscillation conditions is possible.

The circular vibratory conveyor has been designed as an
\lq over-resonance\rq\ apparatus. Low spring constants of the suspensions
set the resonance frequency to $f_{0}\approx 4.5$ Hz. For frequencies
$f  >15$ Hz the change of the amplitude is less than 10 \%. Thus the system
can be assumed as working with constant amplitude independent from the
frequency. However, for execution of accurate experiments the frequency
response of the amplitude
\begin{equation}
A(f)=A_{\infty}\frac{f^{2}}{\sqrt{(f_{0}^{2}-f^{2})^{2}+(2\zeta f_{0} f)^{2}}}
\end{equation}
(with damping $\zeta = 0.08 \pm 0.01$) has been measured. The
dimensionless
acceleration of the conveyor $\Gamma = A(f) \cdot (2\pi f)^{2}/g$, where $g$ is
the gravitational acceleration, can be varied by changing the rotation
frequency $f$ of the unbalanced masses in the range $0 < \Gamma < 7$.

\begin{figure}[tb]
\noindent
\begin{minipage}{74mm}
\epsfxsize=74mm
\epsfbox{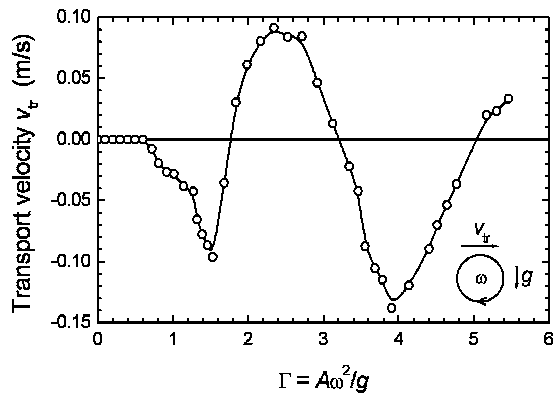}
\caption{Transport velocity $v_{\rm tr}(\Gamma)$ of an object with low
coefficient of restitution (sand-filled \lq fingertip\rq). Conveyor amplitude
$A_{\infty} = 1.7$ mm.}
\label{finger}
\end{minipage}
\end{figure}

For the case of circular vibrations we define the transport direction of the
granulate positive (\lq forward\rq), if the rotational direction of the vibration --
seen from outside the apparatus -- is clockwise while the particles move to the
right, or vice versa. To be more specific, we assign to the circular motion an
axial vector $\vec{\omega}$ (see inset of Fig.\ 2). A positive transport velocity
$\vec{v}_{\rm tr}$ is found, whenever the product
$\vec{v}_{\rm tr}\cdot(\vec{g}\times\vec{\omega})$ is positive.

The granulate used consists of $\approx$ 300 000 glass beads with diameter
$d$ = 1 mm, yielding a layer height of $\approx 5d$. The velocity $v_{\rm tr}$
of the granular flow was measured as a function of the dimensionless
acceleration $\Gamma$ in the range between 0 and 7. The non-monotonous
dependence $v_{\rm tr}(\Gamma)$ is shown in Fig.\ 2. Below a
critical value $\Gamma_{c} \approx 0.45$ the grains stay at rest, i.e., they
follow the agitation of the tray without being transported. The onset of particle
motion is restrained by frictional forces between grains and the substrate. For
accelerations above this threshold the granular material becomes fluidized.
Individual particles are unblocked and begin to move freely on top of each
other. A net granular flow with constant velocity is observed. Note that this
behavior is found already in a regime $\Gamma < 1$ with a vertical
acceleration less than gravity. In order to be transported, the granular
material does not have to leave the ground. It is sufficient to overcome the
frictional forces at the bottom of the granular layer. This transport mechanism
due to stick-slip drag on a horizontally vibrated deck with asymmetric forward
and backward motions is well known to engineers as \lq sliding\rq.

If $\Gamma$ exceeds 1, the vertical component of the circular acceleration
will cause the grains to detach from the bulk followed by a flight on a ballistic
parabola. The corresponding transport mechanism is called \lq throwing\rq.

In our experiments the transport velocity has a first maximum at
$\Gamma = 1.2$. A second maximum is observed at $\Gamma = 4.2$. In
between, the granular flow is slower and even {\em reverses} its direction for
$2.6 < \Gamma < 3.8$.

\begin{figure}[tb]
\noindent
\begin{minipage}{74mm}
\epsfxsize=74mm
\epsfbox{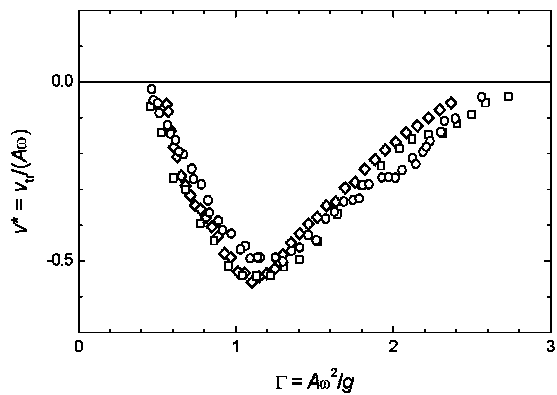}
\caption{Scaled velocity $v^{*} = v_{\rm tr}/(A\omega)$ of the granular flow.
The experimental data were obtained for three different conveyor amplitudes:
$A_{\infty} = 1.0$ mm ($\circ$), $A_{\infty} = 1.7$ mm  ($\square$), and
$A_{\infty} = 2.5$ mm ($\diamond$), respectively.}
\label{scaling}
\end{minipage}
\end{figure}

In contrast, a single 1 mm glass bead is propagated on this vibratory
conveyor in the same direction for all accelerations. The high coefficient of
restitution $\varepsilon \approx 0.9$ for collisions between the glass bead and
the carbon-fiber tray  leads, especially for high $\Gamma$-values, to an
almost \lq random walk\rq\ of the particle in the conveyor with a slight
tendency to propagate in one direction. This is reflected in the large scatter of
the net transport velocity.

In order to study the transport mechanism for a single body in a more
controlled fashion (\lq sandbag test\rq\ \cite{rademacher95}) we designed an
object with a very low coefficient of restitution and non-spherical shape. The
fingertip of a rubber glove was filled with sand. This single object shows
qualitatively the same behavior as the granulate (see Fig.\ 3). The
regime of velocity reversal is, however, shifted to lower values
($1.8 < \Gamma < 3.2$).

By varying the unbalanced mass, the amplitude $A_{\infty}$ of the circular
vibration has been changed between 1.0 and 2.5 mm. We find that the
transport velocity increases linearly with the amplitude. In order to obtain a
dimensionless graph, the granular velocity is scaled by the intrinsic speed of
the driving apparatus, i.e.\ the circular velocity $A\omega$ of the tray. The
resulting master curve (Fig.\ 4) shows clearly that the onset of
particle transport depends only on $\Gamma$. This supports the notion that
the particles have to overcome a (frictional) {\em force}.

An alternative approach would have been to scale the vibratory agitation with
the Froude number $Fr = (A\omega)^{2}/gd = \Gamma A/d$, i.e.\ a typical
dimensionless {\em kinetic energy}. However, the corresponding plot fails to
produce a similar data collapse.

Independent on the oscillation amplitude $A$ we observe that the maximal
achievable transport velocity is equal to $0.5 \cdot A\omega$
(Fig.\ 5). For the single particle with low coefficient of restitution this
is true for both directions of motion, even at higher accelerations. The
reversed granular flow, however, does not attain this limit.

\begin{figure}[tb]
\noindent
\begin{minipage}{74mm}
\epsfxsize=74mm
\epsfbox{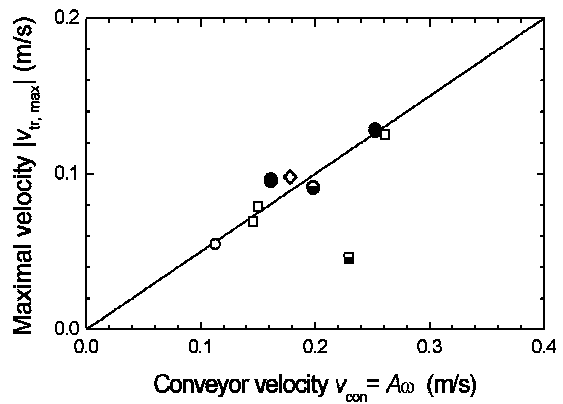}
\caption{Maximal velocity of the granular flow and the fingertip
({\Large$\bullet$}) on the vibratory conveyor. The solid line indicates
$v =0.5 \cdot A\omega$, i.e.\ one half of the local conveyor velocity
$v_{\rm con}$. The experimental data were obtained for three different
conveyor amplitudes: $A_{\infty} = 1.0$ mm ($\circ$), $A_{\infty} = 1.7$ mm
($\square$), and $A_{\infty} = 2.5$ mm ($\diamond$), respectively. Half-filled
symbols indicate reversed flow.}
\label{vmax}
\end{minipage}
\end{figure}

In conclusion, the vibratory conveyor system introduced here opens up the
possibility to investigate the transport properties of granular materials in a
systematic way. Our results show that under certain conditions not even the
direction of the granular flow can be predicted a priori. The delicate
interactions of the particles with the support as well as among themselves
have to be taken into account. First clues indicate that a detailed analysis of
the frictional forces becomes important. A simple model \cite{metcalfe02} for a
frictional block on a horizontally oscillating bed has been introduced recently,
which provides a relaxational mechanism between static and dynamic friction.
This suggests that  before tackling the full problem of many interacting
particles it is even rewarding to investigate the complex behavior of a single
object under the influence of a controlled environment.

We would like to thank H.\ Elhor, S.J.\ Linz, F.\ Landwehr, S.\ Strugholtz, and
T.\ Schnautz for valuable discussions. This work was supported by the
Deutsche Forschungsgemeinschaft (DFG-Sonderprogramm \lq Verhalten
granularer Medien\rq).

\end{document}